\documentclass[]{aa}  
\usepackage{mathtools}
\usepackage{lscape}
\usepackage{longtable,threeparttable}
\usepackage{rotating} 
\usepackage{graphicx}
\usepackage{multicol}
\usepackage{wasysym}
\usepackage{tablefootnote}

\usepackage{txfonts,booktabs}
\usepackage{hyperref}
\hypersetup{
    unicode=false,
    pdftoolbar=true,
    pdfmenubar=true,
    pdffitwindow=false,
    pdfstartview={FitH},
    pdftitle={My title},
    pdfauthor={Author},
    pdfsubject={Subject},
    pdfcreator={Creator},
    pdfproducer={Producer},
    pdfkeywords={keyword1} {key2} {key3},
    pdfnewwindow=true,
    colorlinks=true,
    linkcolor=red,
    citecolor=blue,
    filecolor=magenta,
    urlcolor=cyan
}

\begin{document}
   \title{The MUCHFUSS photometric campaign}

   \titlerunning{Photometric follow-up of the MUCHFUSS project}
   \authorrunning{Schaffenroth et al.}

   \author{V. Schaffenroth \inst{1,2,3} \and  S.Geier \inst{1,4} \and U. Heber\inst{3} \and R. Gerber \inst{3} \and D. Schneider\inst{3} \and E. Ziegerer \inst{3} \and O. Cordes \inst{5}}
   \institute{Institute for Astronomy and Astrophysics, Kepler Center for Astro and Particle Physics, Eberhard Karls University, Sand 1, 72076,
         T\"ubingen, Germany \email{schaffenroth@astro.uni-tuebingen.de}
         \and Institut für Astro- und Teilchenphysik, Universität Innsbruck, Technikerstr. 25/8, 6020 Innsbruck, Austria\\       
         \and Dr.\,Remeis-Observatory \& ECAP, Astronomical Institute, Friedrich-Alexander University Erlangen-N\"urnberg, Sternwartstr.~7, 96049 Bamberg, Germany
         \and Department of Physics, University of Warwick, Coventry CV4 7AL, UK
         \and Argelander Institute for Astronomy, Auf dem H\"ugel 71, D-53121 Bonn, Germany
             }

   \date{Received 26 September 2016/ Accepted 9 February 2018}

\abstract{
Hot subdwarfs (sdO/Bs) are the helium-burning cores of red giants, which lost almost all of their hydrogen envelopes. This mass loss is often triggered by common envelope
interactions with close stellar or even substellar companions. 
Cool companions like late-type stars or brown dwarfs are
detectable via characteristic light curve variations like reflection effects and often also eclipses. To search for such objects 
we obtained multi-band light curves of 26 close sdO/B binary candidates from the MUCHFUSS project with the BUSCA instrument. We discovered a new eclipsing reflection effect system ($P=0.168938$~d) with a low-mass M dwarf companion ($0.116 M_{\rm \odot}$). Three more reflection effect binaries found in the course of the campaign were already published, two of them are eclipsing systems, in one system only showing the reflection effect but no eclipses the sdB primary is found to be pulsating. Amongst the targets without reflection effect a new long-period sdB pulsator was discovered and irregular light variations were found in two sdO stars.
The found light variations allowed us  to constrain the fraction of reflection effect binaries and the substellar companion fraction around sdB stars.
The minimum fraction of reflection effect systems amongst the close sdB binaries might be greater than 15\% and the fraction of close substellar
companions in sdB binaries  might be as high as $8.0\%$. This would result in a close substellar companion fraction to sdB stars of about 3\%. This fraction is much higher than the fraction of brown dwarfs around possible progenitor systems, which are solar-type stars with substellar companions around 1 AU, as well as close binary  white dwarfs with brown dwarf companions. This might be a hint that common envelope interactions with substellar objects are preferentially followed by a hot subdwarf phase.
}

   \maketitle

\section{Introduction}

Stars do not form in isolation and therefore most of them are members of binary or multiple systems \citep[see][for a review]{duchene}. The separations of a significant fraction of those binary pairs are so small, that the stars will interact with each other during their lifetimes. Several studies determined the frequency of sufficiently close companions around solar-type stars \citep[see][and references therein]{winn:2015,sahlmann:2011}. The fraction of stellar companions closer than 10 AU is about 13\%. 

While interactions in the main sequence phase are possible for the closest systems, it is more likely that the stars start to interact in the post-main sequence phase as soon as the more massive star expands and evolves to become a red giant increasing its radius by two to three orders of magnitude. Depending mostly on the mass ratio the resulting interaction will lead to mass-transfer, which can be either stable or unstable. In the most extreme cases, this interaction can change the evolution of the stars dramatically and form objects, that can hardly be explained otherwise \citep[see][Chapter 2, for a review]{binary}. 

An important class of such objects are the hot subdwarf stars (sdO/Bs), which are evolved, core helium-burning objects with thin hydrogen dominated envelopes and masses around $0.5\,M_{\rm \odot}$ \citep[see][for a review]{heber:2009,heber:2016}. The progenitor star has to lose almost its entire hydrogen envelope in the red-giant phase to form such an object. About half of the sdB stars reside in close binaries with short periods from just about one hour to a few days \citep{maxted:2001}. Such close binaries are believed to be formed by a common-envelope and spiral-in phase \citep{han:2002,han:2003}, as the separation in these systems is much smaller than the size of the red-giant progenitor star. In this scenario of unstable mass-transfer, a low-mass companion is swallowed by the close-by red giant. Subsequently, the companion spirals towards the core of the red giant while removing the hydrogen envelope. Although the common-envelope ejection channel is not properly understood in detail \citep[see][for a review]{ivanova}, it provides a reasonable explanation for the strong mass loss required to form sdB stars. Such close post-CE binaries (PCEBs) are also found among the white dwarf population. However, since the majority of stars end their lives as white dwarfs, the relative fraction of PCEBs is much smaller than for the sdOB stars. 

It is known for decades that the lowest mass stars close to the hydrogen burning limit can expel a common envelope \citep[e.g.,][]{aador_kilkenny}. 
Substellar objects with masses below that limit have been found in close orbits around stars as well. About 10\% of the solar-type stars have giant planets with masses greater than $0.1\,M_{\rm Jup}$ in orbits  smaller than 10 AU. 
The more massive brown dwarf (BD) companions on the other hand are found only around $<1\%$ of the solar-type stars. This paucity of BDs is also known as the brown dwarf desert.
All those close substellar companions will eventually interact with their host stars.
The discovery of three sdB systems with close brown dwarf companions \citep{geier,vs:2014_I, vs:2015a} provides the best evidence that substellar objects are able to interact with a star in a common envelope phase and form an sdB star.
 
\citep{han:2002,han:2003} performed a binary population study to predict the fractions of hot subdwarf stars coming from different formation channels using Monte-Carlo simulations with different model parameters. Their favoured model predicts a fraction of 18\% of the hot subdwarf stars coming from the common envelope channel, which produces hot subdwarf stars with low-mass stellar or substellar companions, most of the companions being M dwarfs. Acounting for selection effects they predict that the measurable fraction of low-mass stellar or substellar companions to hot subdwarf stars should be  about 40\%, when the selection is sensitive to short-period binaries like in \citet{maxted:2001}, for example. 
	
 Constraining the fraction of low-mass stellar and substellar companions to hot subdwarf stars can hence help for a better understanding of the formation of hot subdwarf stars and the common envelope phase. The current sample of hot subdwarf binaries consists of 142 binaries, 26 of them with low-mass stellar or substellar companions \citep{kupfer:2015}, which relates to a fraction of 15\%. However, the selection is quite inhomogeneous and is therefore not suited to derive the fraction of low-mass stellar or substellar companions to hot subdwarf stars. 

Several different typical light variations are found in sdB binaries (more details in Sect. \ref{variations}) allowing to restrict the nature Here we present
the photometric follow-up of targets from the MUCHFUSS project. 
In Sect. \ref{analysis}
we describe the observations and present the already published results of our photometric follow-up campaign. In Sect. \ref{new} we present the new discovery of an eclipsing sdB binary with a low mass M dwarf companion and the discovery of a new sdB pulsator in a binary. Moreover, we show two He sdOs with light variations. In Sect. \ref{substellar} we try to confine the fraction of substellar companions.
Finally, in Sect. \ref{discussion} we discuss our results.

\section{The MUCHFUSS photometric follow-up campaign}
\label{analysis}

The MUCHFUSS (Massive Unseen Companions to Hot Faint Stars from SDSS) project was initially designed to find massive compact companions like massive white dwarfs, neutron stars or even black holes as companions to hot subdwarf stars. To discover rare objects, a huge initial dataset is necessary. The enormous SDSS database was therefore the starting point for the survey. Hot subdwarfs are found most easily by applying a colour cut to Sloan photometry. The spectra of the colour-selected stars were subsequently classified by visual inspection. 

Since the spectra of hot subdwarfs in close binary systems do not show any lines of the companions, we then selected all single-lined sdO/Bs with spectra that are sufficiently bright ($V < 18.5$ mag) to have spectra with reasonable S/N ($>5$). The SDSS spectra are coadded from three individual spectra with an exposure time of $15\,{\rm min}$ taken consecutively. Measuring the radial velocities of those individual spectra and performing a large spectroscopic follow-up campaign, we selected close binary sdB stars, which show significant RV-variability \citep{geier:2011_2,geier:2015,geier:2017} for the photometric follow-up. In addition to this thoroughly selected sample (see Table~\ref{mass}) we also observed several hot {sub\-dwarf} stars as backup targets (see Table~\ref{backup},\ref{sdo}).
This selection turned out to be not only well suited to find compact companions, but also low-mass main-sequence or substellar companions with very short orbital periods \citep[e.g.,][]{geier:2011,vs:2014_I}.

\subsection{Light variations found in sdB binaries}\label{variations}
\begin{figure}
\centering
\includegraphics[width=1.0\linewidth]{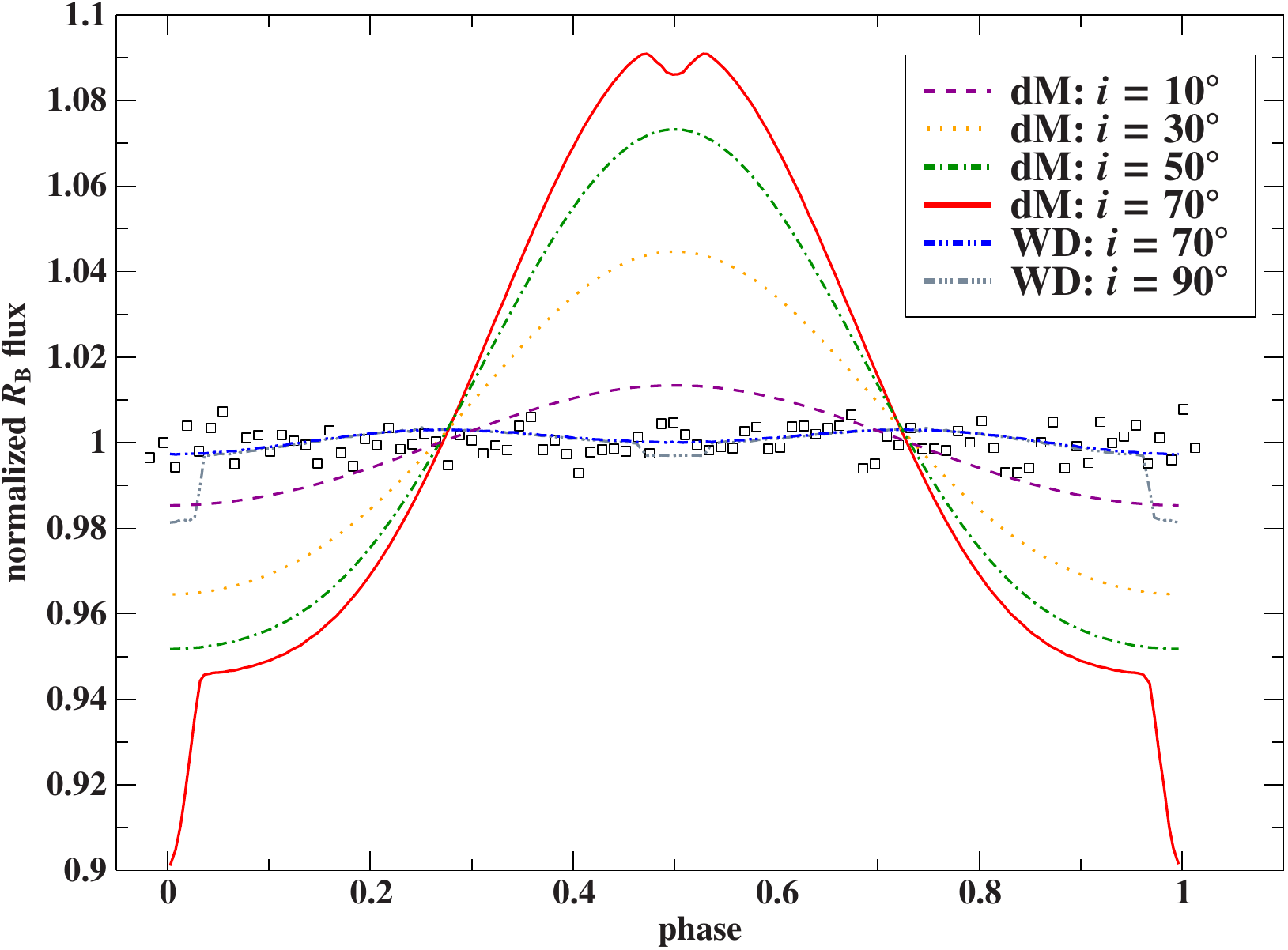}
\caption{Different lightcurve models showing the typical light variations of hot subdwarf binaries superimposed by a typical non-detection lightcurve of the photometric Muchfuss follow-up. All lightcurve models were calculated with the MORO code \citep{drechsel:1995} in photometric band $R_{\rm B}$. We calculated models of a sdB+dM system with the parameters of HW Vir (see Table \ref{HW Vir}) representing a typical HW Virginis system at different inclinations showing the reflection effect and of a sdB binary system with a low mass WD with the same period and RV semi-amplitude at 90$^\circ$ and 70$^\circ$. Variations in the lightcurve of the sdB+WD systems are due to ellipsoidal deformation of the sdB.}
\label{hw_vir}
\end{figure}

In our spectroscopic follow-up campaign we measured the radial velocity curves and orbital parameters of the binaries. However, in such single-lined spectroscopic binaries only minimum masses can be derived for the companions, as the inclination is unknown. To unravel the nature of the companions, photometric follow-up is helpful, because the diverse classes of companions can cause characteristic signatures in the binary lightcurves. 

The so-called reflection effect is observed in close binary systems consisting of a hot primary and a cool companion. As the secondaries in these systems
are supposed to orbit synchronously, the hemisphere of the cool companion facing the hot primary is constantly irradiated and heated up, which leads to an increased flux around the secondary eclipse \citep[see][for a detailed discussion of this effect]{wilson:1990,budaj}.
The amplitude of the reflection effect scales with the temperature ratio and the radii of the primary and secondary star, as well as the inverse orbital separation \citep{wilson:1990,budaj}.
Hence, the reflection effect is strong when both components of a close binary system have a very small separation, similar radii, and a high temperature difference. In systems consisting of a low-mass main sequence star or a brown dwarf and a hot, compact star like a hot subdwarf, these conditions are fulfilled. 
The reflection effect is visible as a sinusoidal variation with a period equal to the orbital period (see Fig. \ref{hw_vir}).

Due to the short periods of such sdB binaries these systems also have a high probability to be eclipsing. Eclipsing systems are of high value because they allow to determine the masses and radii of both components with a combined photometric and spectroscopic analysis. Moreover, the separation of the system can be determined. Eclipsing binaries with low-mass stellar or substellar companions are called HW Virginis (HW\,Vir) systems after their prototype. Only 19 of this systems are published so far, which makes this type of system quite rare. Due to the unique shape of the lightcurve almost all of them have been found in photometric surveys or by chance while looking for pulsations.

In Fig.~\ref{hw_vir} the dependency of the reflection effect amplitude on the inclination angle is illustrated with synthetic lightcurves adopting the parameters of the prototype HW Vir, which represents a typical HW Vir system (see Table \ref{HW Vir} and Fig. \ref{logP}). Even for an inclination angle as small as 10$^\circ$ variations of almost 2\% are predicted, which are detectable from the ground. Multi-band lightcurves are very useful to confirm the reflection effect. Since the temperature of the irradiated hemisphere of the companion is heated up to a temperature of about 10\,000-20\,000~K \citep{vuckovic:2016}, which is lower than the temperature of the sdB, the observed variation is wavelength dependent and becomes stronger at longer wavelengths, where the contrast in brightness between the irradiated companion and the hot subdwarf becomes stronger. Due to the much smaller radius of the companion the reflection effect is not visible in sdB+WD systems (see Fig. \ref{hw_vir}).

However, in the shortest-period sdB binary systems with WD companions the tidal influence of the rather massive and compact companion can lead to an ellipsoidal deformation of the sdB. This results in a sinusoidal variation with half the orbital period \citep[e.g.,][]{wilson:1997}. The amplitude of the ellipsoidal variation scales with the mass of the companion and the inverse orbital period. 
Since it is caused by the distorted shape of the primary, the wavelength dependence  of the variation is much less than for the reflection effect.

Multi-band photometry, therefore, allows us to distinguish between variations due to the reflection effect or due to ellipsoidal deformation. Morover, Fig. \ref{hw_vir} shows that the reflection effect in sdB binaries is much stronger than the ellipsoidal deformation for a given orbital period.

In some cases multiperiodic light variations due to pulsations of the sdB primary are detected in addition to the binary signatures. Two main types of sdB pulsators are known. \citet{kilkenny:1997} discovered the ${\rm sdBV}_r$ stars, also called V361 Hya stars, which show non-radial p-mode pulsations with short periods of $\sim100-300\,\rm s$ and amplitudes from a few ppm to about 5 \%. 
The second class was found by \citet{green:2003}. These stars, called ${\rm sdBV}_s$ stars or V1093 Her stars, show high radial-order g-mode pulsations with periods of $\sim 2000-8000\,\rm s$ and very small amplitudes below 0.5 \%. 
Since the pulsation periods of the V1093 Her stars are very similar to the orbital periods of the closest sdB binaries and can show a complex wavelength dependency, such pulsations can be misclassified as shallow reflection effects or ellipsoidal variations. 
 
Recently, apparently irregular light variations have been discovered in some sdO stars, the hotter siblings of the sdBs. It is not yet clear where these strange variations come from, but they look  quite similar to variations found in cataclysmic variables \citep{tuc_betsy}.

\subsection{Observations}\label{observations}
The number of known hot subdwarf stars with cool low-mass companions (reflection effect systems) is still small and most of them have been discovered due to their characteristic light curves in surveys looking for pulsating stars or from the data archives of planetary transit and transient surveys \citep[e.g.,][]{vs,almeida:2012}.

However, the fraction of low-mass stellar and substellar companions in hot subdwarf binaries cannot be derived based on this heterogeneous sample. This motivated us to conduct a photometric follow-up of 26 spectroscopically selected sdB binary candidates from the MUCHFUSS project \citet{geier:2015,geier:2017}. In addition to the selection criteria described in Sect.~2, we picked systems brighter than $g\simeq18\,{\rm mag}$ to obtain light curves with an S/N sufficient to find also small light variations. Moreover, we preferentially chose systems which already got additional spectroscopic follow-up and therefore highly significant RV variations. If no such systems were observable at a given time backup targets were chosen based on their visibility and magnitude depending on the observing conditions.

\begin{figure}
	\centering
	\includegraphics[width=1.0\linewidth]{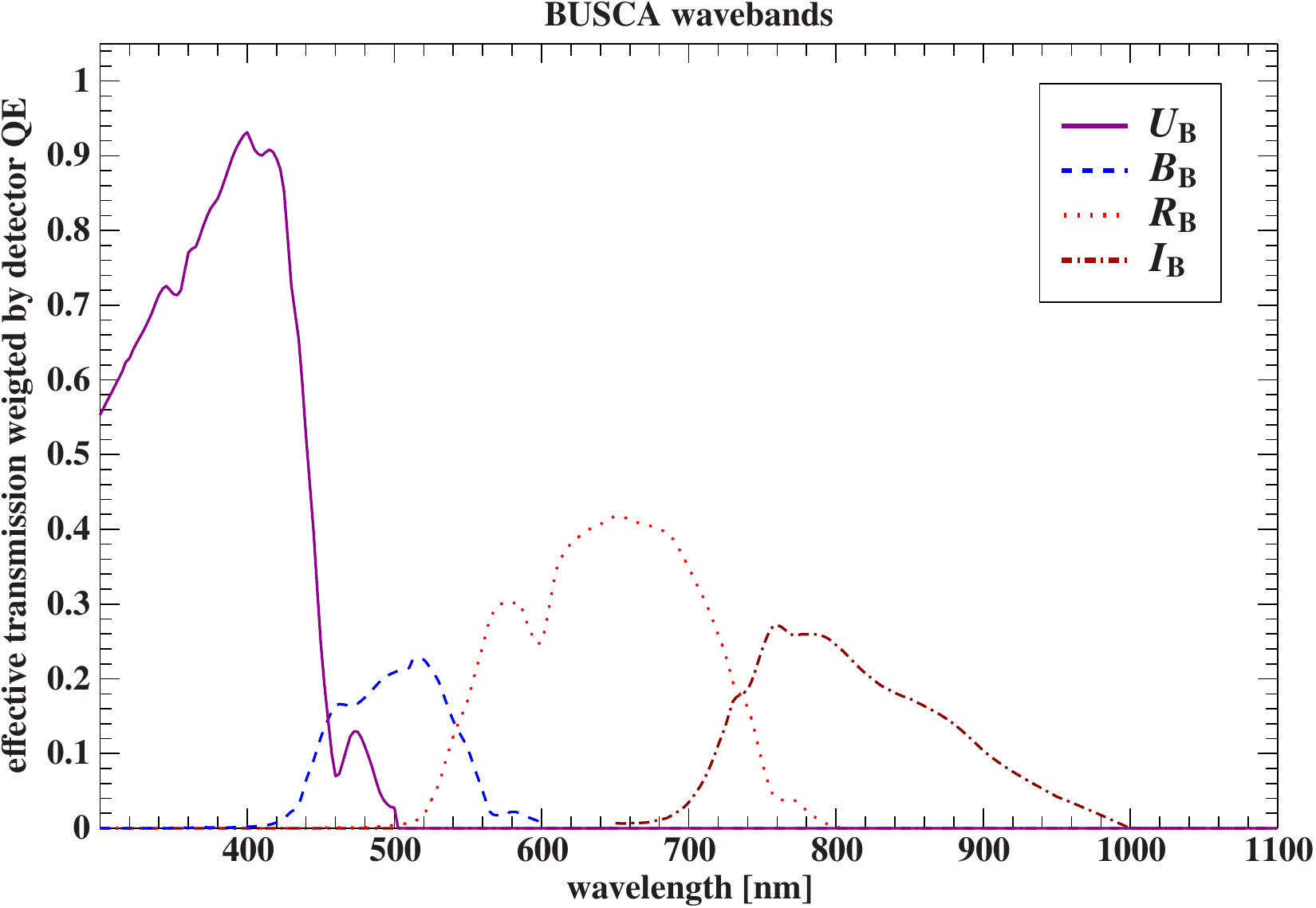}
	\caption{Transmission curves of the BUSCA passbands taken from the BUSCA homepage (\url{http://www.caha.es/CAHA/Instruments/BUSCA/bands.txt}) weighted by the quantum efficiency of the detector (\url{https://www.caha.es/guijarro/BUSCA/QE_thick_thin.txt}).}
	\label{busca}
\end{figure}

This follow-up was done mostly with BUSCA \citep[Bonn University Simultaneous CAmera; see][]{reif} on the 2.2m-telescope located at the Calar Alto Observatory in Spain. This instrument turned out to be perfect for our purposes, as it is possible to observe in four bands simultaneously. We did not use any filters but the intrinsic transmission curve given by the beam splitters ($U_B$, $B_B$, $R_B$, $I_B$). In this way all visible light is used most efficiently. The transmission curves of the BUSCA bandpasses are shown in Fig. \ref{busca}.

\begin{table}\centering
\caption{Log of observations with BUSCA}
\begin{tabular}{c c c}\hline\hline
Year/Month/Date&P.I&Observer\\\hline
2010/06/11&SG&VS\\
2010/09/29-2010/10/03&SG&VS \& EZ\\
2011/02/25-2010/03/01&SG&VS\\
2011/05/30,31;2011/06/03&OC&VS\\
2011/09/28-2011/10/02&SG&RG\\
2011/10/17-21&OC&VS\\
2012/01/13,14&SG&EZ\\
2012/10/10-14&SG&VS\\

\hline
\end{tabular}
\label{observation}
\tablefoot{\\
VS (V. Schaffenroth); SG (S. Geier); EZ (E. Ziegerer); OC (O. Cordes); RG (R. Gerber)}
\end{table}

The data was taken during several observing runs listed in Table \ref{observation}. In total, 30 nights were granted for this programme with BUSCA. For each star time-resolved photometry with a duration between 1.5 and 2.5 h and exposure times from 30 to 180 seconds per image was taken. The strategy was tailored to find light variations with periods of 2 to 3 hours, which matches the typical orbital periods of the HW Vir systems.

Defining smaller windows around the stars within the full images, the read-out time was reduced from 2 min for the full-frame image to about 15-20 seconds. We observed our target and four comparison stars in 60 x 60 pixel windows to perform differential photometry. The comparison stars were selected using the SDSS Navigate Tool \footnote{\url{http://skyserver.sdss.org/dr10/en/tools/chart/navi.aspx}}. We chose comparison stars with similar magnitudes ($\Delta m < 2$ mag) in all SDSS bands from u to z.
The data reduction was done using IRAF\footnote{\url{http://iraf.noao.edu/}}. The standard CCD reduction was done using the tools for bias-, and flatfield-correction and the lightcurves were extracted using the aperture photometry package {\sc daophot}. To correct for changing airmass and conditions the lightcurve of the target was divided by the lightcurves of the comparison stars.

\label{observ}
\subsection{Summary of published light variations}
\label{1622}

In the course of our photometric follow-up campaign we already discovered and published several objects. The spectra of SDSS J082053.53+000843.4 (J0820) and SDSS J162256.66+473051.1 (J1622) showed RV variations with short periods and the photometric follow-up confirmed them to be eclipsing with a period of 0.096 d for J0820 \citep{geier}\footnote{In this case the lightcurve was taken with the Merope instrument mounted at the 1.3m-Mercator telescope at the Roque de los Muchachos observatory on La Palma} and 0.069 d for J1622 \citep{vs:2014_I}. The companions of both systems have masses of 0.045 to 0.068 $M_{\rm \odot}$ and 0.064 $M_{\rm \odot}$, respectively. Those systems were the first confirmed sdB+BD systems. 

Finally, we discovered the sdB binary FBS\,0117+396 showing the reflection effect as well as pulsations. In addition to two short-period pulsation modes, we detected several modes with long periods. FBS\,0117+396 is therefore the first hybrid pulsator in a reflection effect binary with an M-dwarf companion \citep{ostenson:2013}.

\section{New discoveries}
\subsection{J192059+372220 -- a new HW Vir system}
\label{new}
\begin{figure}
	\includegraphics[angle=-90,width=1.0\linewidth]{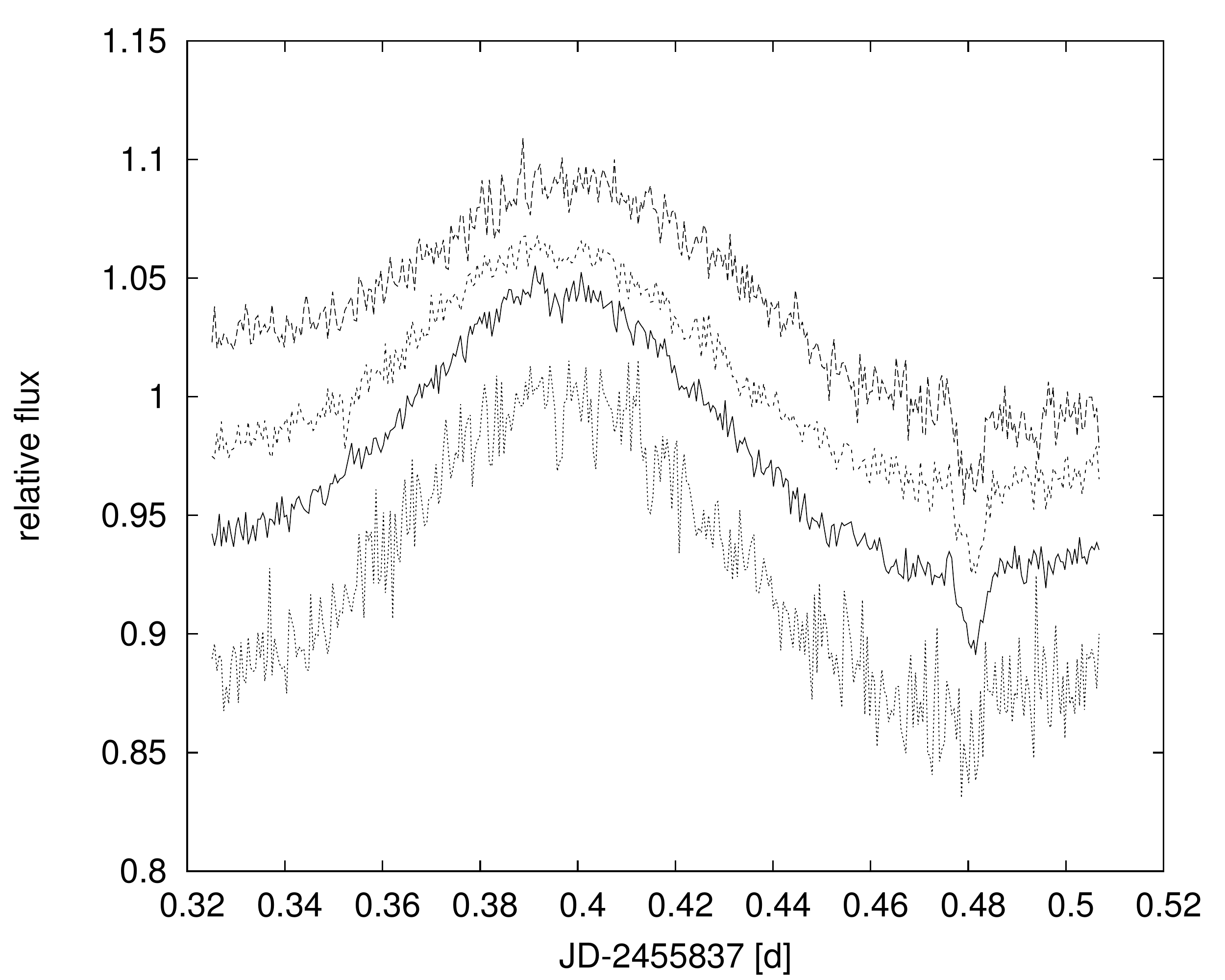}
	\caption{BUSCA lightcurve of the eclipsing sdB+dM system J1920 ($U_B,B_B,R_B,I_B$, from top to bottom).}
	\label{1920}
\end{figure}
\begin{figure}
\centering
\includegraphics[width=1.0\linewidth]{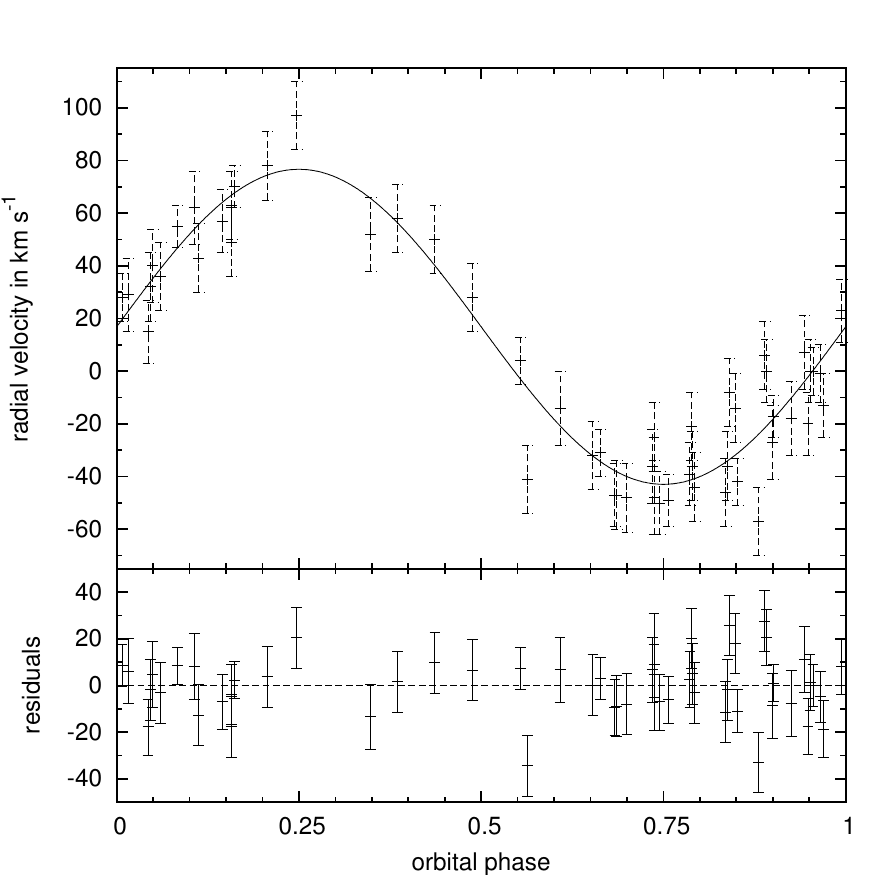}
\caption{Radial velocity curve of J1920 with the best fit superimposed. The residuals are shown in the lower panel.}
\label{rv}
\end{figure}

During the course of our photometric follow-up with BUSCA
we discovered the new HW Vir system J192059+372220 (J1920), which was selected as RV variable sdB. The observations were taken and reduced as described in Sect. \ref{observations} during the two runs in Sept/Oct 2010/2011. Its lightcurve, which is shown in Fig.~\ref{1920}, shows a prominent reflection effect and grazing eclipses. We obtained 39 medium resolution (grism T13, $R\sim4000$, wavelength coverage $3450 - 5750\,\AA$) spectra with the TWIN spectrograph on 26/27/29/30 May 2012 at the 3.5m telescope at the Calar Alto observatory, which were reduced with the MIDAS package\footnote{\url{http://www.eso.org/sci/software/esomidas//}. Furthermore, 11 spectra with grism R600 giving similar resolution (coverage from 3500 to 5300 $\AA$) from the ISIS spectrograph at the William Herschel Telescope at the Roque de los Muchachos Observatory on La Palma were taken on the 9 and 11 July 2012, which were reduced using IRAF$^2$. The exposure time of the TWIN and ISIS spectra was 20 min.}

Those spectra were used to determine the radial velocity curve. The RV was measured using the SPAS routine \citep{hirsch} by fitting Gaussians and Lorentzians to the Balmer and helium lines. A circular orbit was assumed due to the short period and, therefore, a sine curve was fitted to the measured RVs to determine the RV semi-amplitude $K=57.2\pm 2.3\,\rm km\,s^{-1}$, the system velocity $\gamma=18.0\pm1.7 \,\rm km\,s^{-1}$, and the orbital period $0.168938\pm0.00011\,{\rm d}$. The results are summarized in Table \ref{result}.

The atmospheric parameters of the sdB primary were determined by fitting synthetic spectra, which were calculated using LTE model atmospheres with solar metallicity and metal line blanketing \citep{heber:2000}, to the Balmer and helium lines. Each spectrum was fitted separately, because in several other HW Vir systems an apparent change of the atmospheric parameters with the orbital phase has been detected \citep{vs,heber:reflection}, which is caused by the time-variable contribution of the irradiated companion. This effect is only visible when the atmospheric parameters can be derived with sufficient accuracy. In our analysis this effect was not visible and thus we co-added all spectra, which were shifted to RV zero before, to improve the S/N. The results are summarized in  Table \ref{result}. The effective temperature $T_{\rm eff}=27500\pm 1000 \rm\, K$ and the surface gravity of $\log{g}=5.4\pm0.1$ are typical for an sdB situated on the extreme horizontal branch.

\begin{table}\centering
\caption{Atmospheric and fundamental parameters of J1920.}
\begin{tabular}{c| l l c l}\hline\hline
    coord.&ep=2000&\multicolumn{3}{l}{19 20 59 +37 22 20}\\
    g$'$&[mag]&15.58&\\
	\hline
	K & $\rm [km\,s^{-1}]$ & 57.2 & $\pm$ & 2.3\\
	$\gamma$ & $\rm [km\,s^{-1}]$ &18.0  & $\pm$ & 1.7\\
	P & [d] & 0.168938 & $\pm$ & 0.00011\\
	a & [$\rm R_{\odot}$] & 1.078 & $\pm$ & 0.0449\\
	$T_{\rm eff}$ & [K]& 27500& $\pm$ & 1000\\
	$\log{g}$ &[cgs] & 5.4&$\pm$&0.1\\
	$\log{y}$&[cgs]&-2.5&$\pm$ & 0.25\\
	i & [$^\circ$] & 67 & $\pm$ &2 \\
	$M_{\rm sdB}$ & [$M_{\rm \odot}$]&0.47&\\
	$M_{\rm comp}$ & [$M_{\rm \odot}$]&0.116 & $\pm$ & 0.007\\
	\hline
\end{tabular}
\label{result}
\end{table}

From the BUSCA lightcurves we could determine the orbital period to be 0.168965(43) d, which is consistent with the period determined from the RV curve. Unfortunately, the S/N of our data is not sufficent for a proper lightcurve analysis. A preliminary analysis using MORO \citep[MOdified ROche program, see][]{drechsel:1995} as described in e.g., \citet{vs:2014_I} indicates that the inclination angle of the system is about $67^\circ$. Adopting this value and assuming a canonical mass of 0.47 $M_{\rm \odot}$ for the sdB, we derive a companion mass of 0.116 $M_{\rm \odot}$ from the binary mass function. This mass estimate is consistent with a late M dwarf making J1920 a quite typical HW\,Vir system.

\subsection{GALEX\,J050735+034815 -- A newly discovered pulsator}
\begin{figure}
	\includegraphics[angle=-90,width=1.0\linewidth]{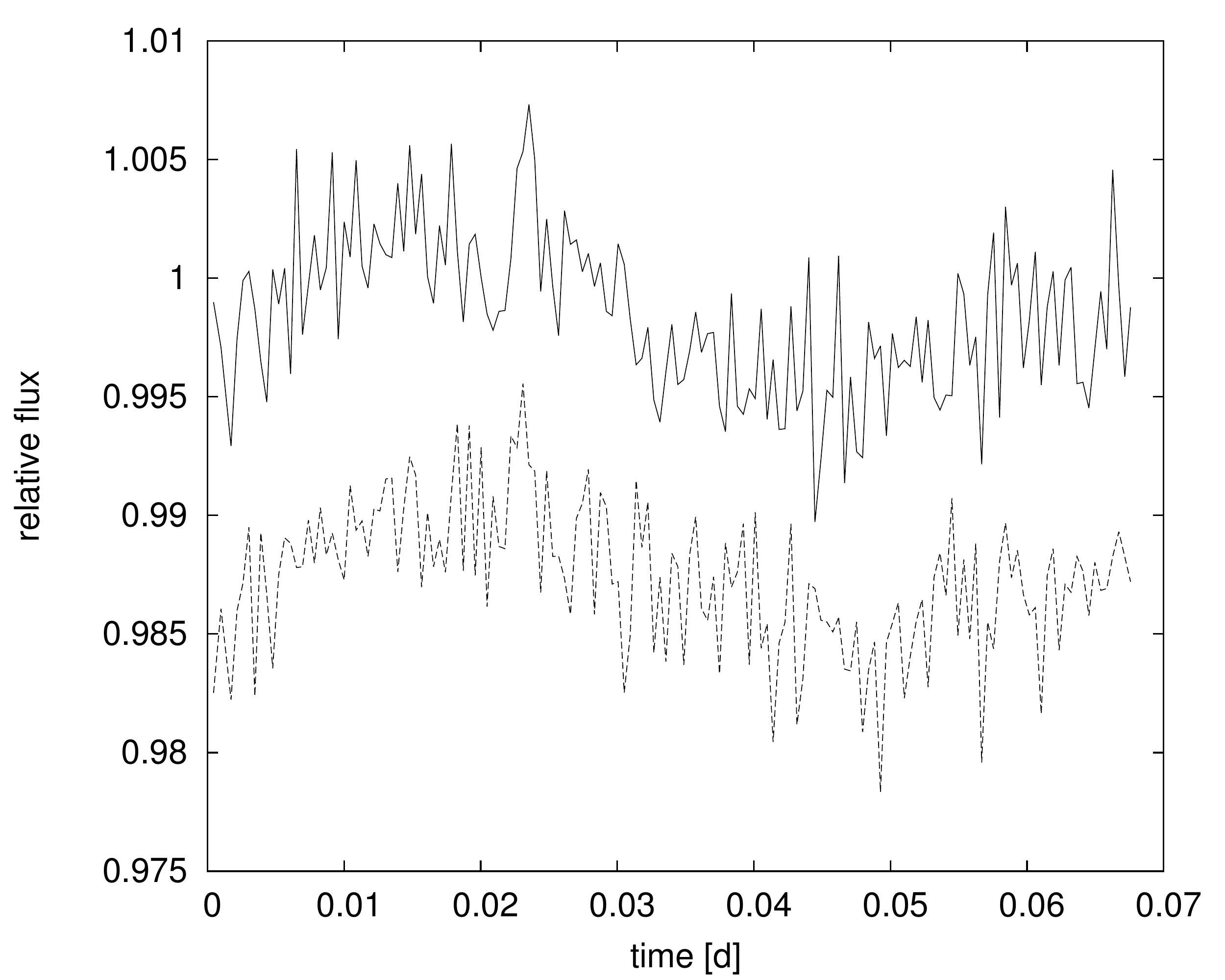}
	\caption{BUSCA lightcurve of J0507 in $B_B$ and $R_B$ showing most probable long-period g-mode pulsations.}\label{J0507}
\end{figure}
For the bright backup target\footnote{see Table \ref{backup} for a list of observed backup targets} GALEX\,J050735+034815 (J0507) \citep{nemeth:2012} we detected a small sinusoidal variation with a period of about 2.7~h, which is shown in Fig. \ref{J0507}. Since the amplitude of the variation is similar in the $V_B$ and $R_B$ band a reflection effect can be excluded. Initially, we suspected the variation to be caused by an ellipsoidal deformation and the sdB to be in a close binary with a WD companion similar to the sdB+WD binary CD$-$30$^\circ$11223, which was also found in the MUCHFUSS project \citep{cd-30}. 

However, most recently \citet{kawka:2015} determined the binary parameters of this system based on spectroscopy. The orbital period of $\sim0.528\,{\rm d}$ is much longer than the period of the light variation we detected, which can therefore not be an ellipsoidal variation. \citet{nemeth:2012} determined the atmospheric parameters of the sdB star. With an effective temperature of $T_{\rm eff}=23990\,{\rm K}$ and a surface gravity of $\log{g}=5.42$ the star is situated in the instability region of the long-period sdB pulsators. Since period and amplitude of the observed variation are consistent with a pulsation mode, we conclude that J0507 is most likely an V1093 Her pulsator.  

\subsection {Variations of He-sdOs}
\label{var_sdo}
\begin{figure}
 \includegraphics[angle=-90,width=1.0\linewidth]{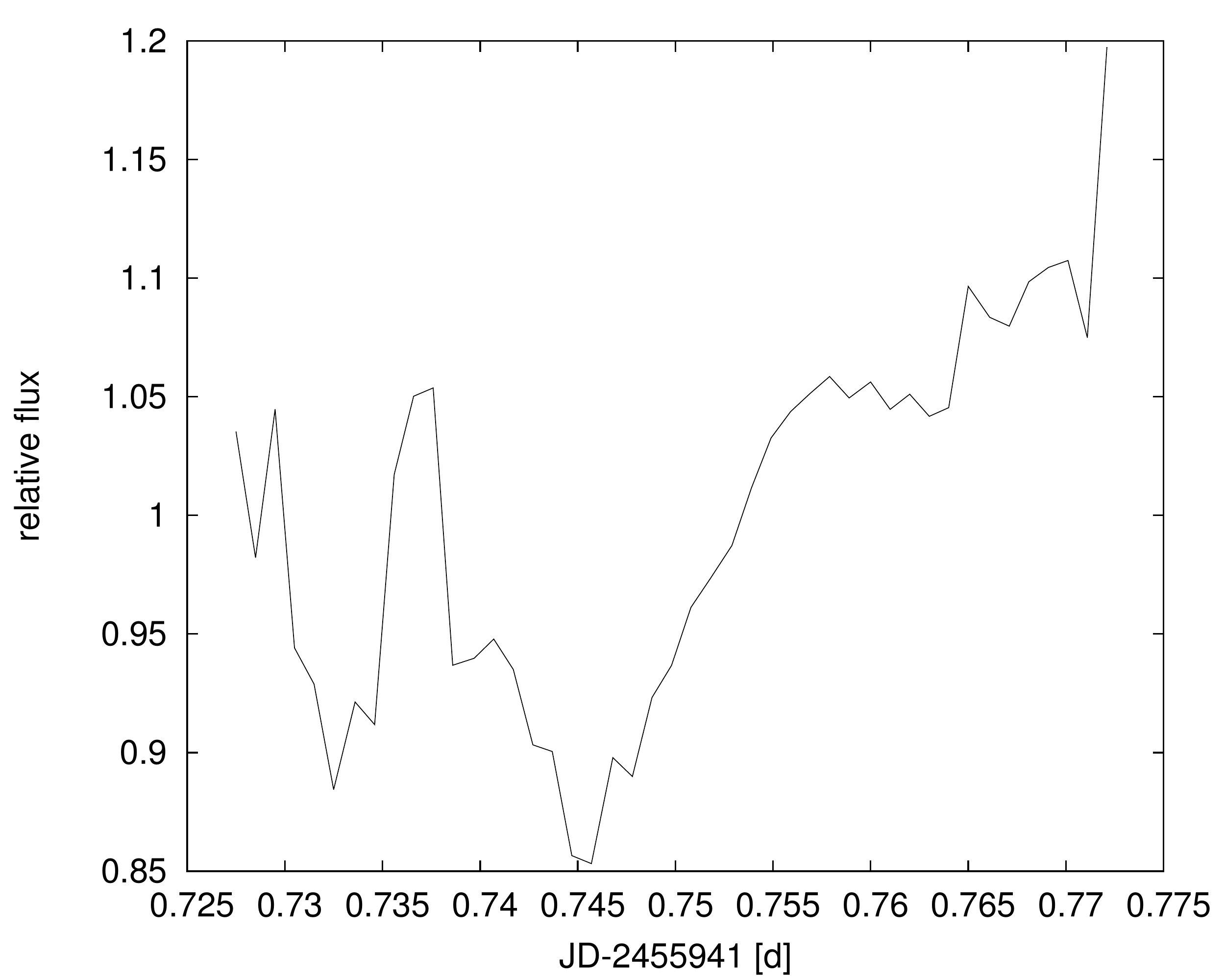}
 \caption{BUSCA lightcurve of He-sdO J1415 in $R_B$ showing strong irregular variations.}
 \label{hesdo}
 \end{figure}
  \begin{figure}
  \includegraphics[angle=-90,width=1.0\linewidth]{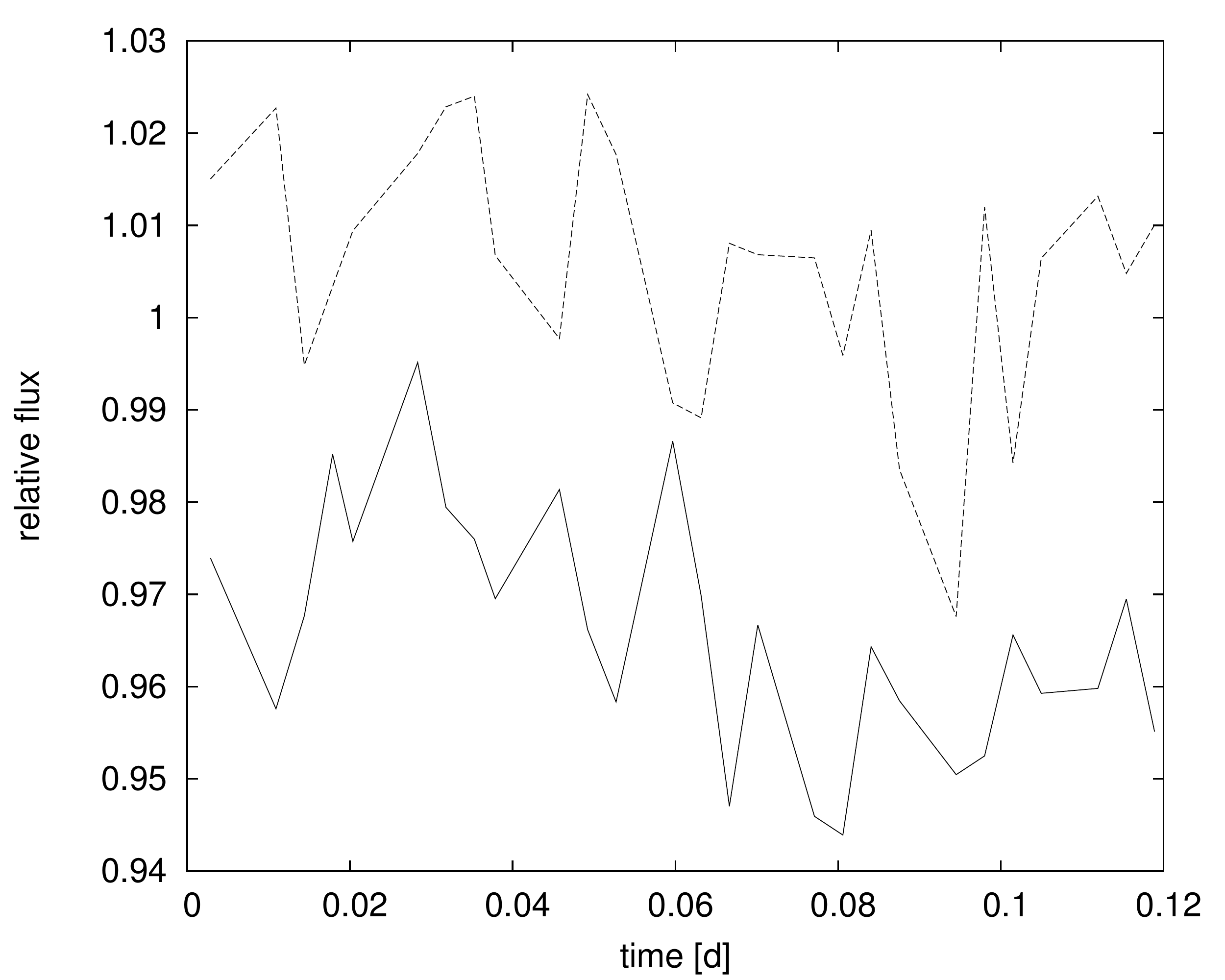}
  \caption{BUSCA lightcurve of the He-sdO J1124 showing low-amplitude light variations ($B_B$ and $R_B$).}
  \label{J1124}
  \end{figure}
 
As already shown in \cite{tuc_betsy} and \cite{geier:2015} several helium-rich sdO stars have been found to show irregular RV and light variations. To investigate this further we observed several He-sdOs. J141549+111213 (J1415) shows irregular variations \citep[see Fig. \ref{hesdo} and][]{geier:2015}. The origin of these variations is still unclear. Moreover, the He-sdO J112414+402637 (J1124) shows a low-amplitude sinusoidal variation (see Fig. \ref{J1124}). The cause of this variation is also unclear and further RV and lightcurve observations are necessary to investigate these systems. No significant variations have been detected in seven other sdO and He-sdO stars (see Table \ref{sdo}). 

\section{The fraction of low-mass stellar and substellar companions}
\label{substellar}

As shown before four out of the 26 observed sdB or sdOB binaries with significant RV shifts show a reflection effect. Three of them show eclipses and two of them have substellar companions. For 21 of the observed sdB or sdOB systems, it is possible to exclude typical sdB+dM systems with parameters similar to HW Vir and also sdB+BD systems with parameters similar to J1622 applying the method outlined in Appendix~\ref{method}. One of the lightcurves is too noisy and/or the measured RV shifts are too small to be conclusive. In this case, a low-mass stellar and substellar companions cannot be firmly excluded. 

Therefore, we conclude that 81\% of our sample are not HW\,Vir-type systems, 15\% are HW\,Vir or reflection effect systems, 8\% of the sdB binaries of our sample have substellar companions and for another 4\% no conclusions can be drawn. Since we can only firmly exclude systems with short orbital periods, the fractions of low-mass stellar and substellar companions have to be regarded as lower limits only. 

Although HW\,Vir systems have been studied since decades, no other sample allows us to constrain their fraction of reflection effect and HW\,Vir binaries  in a similar way. The discovery history of those systems (see Table \ref{HW Vir}) is very inhomogeneous. They stick out due to their characteristic light curves. Some are already known for decades (e.g. AA\,Dor, HW\,Vir), others have been found in surveys for pulsating sdB stars (e.g. NY\,Vir, HS\,2231+2441, HS\,0705+6700, EC\,10246-2707). The most recent discoveries are from large light curve archives of transit (e.g. ASAS\,10232, NSVS\,14256825), microlensing (e.g. BUL$-$SC16\,335) or transient (e.g. PTF1\,J072455.75+125300.3) surveys. This sample can therefore not be used to draw any quantitative conclusions.

The most comprehensive dataset of sdO/B light curves to date is the one collected by \O stensen et al. searching for pulsations \citep[e.g.][]{oestensen_pulsators}. The most recent catalog contains 776 sdO/Bs with light curve data (\O stensen priv. comm.). Only 12 systems of the MUCHFUSS target list are included, because the SDSS sample is fainter than the rest. Four of them were also observed by us with BUSCA. But more importantly, those light curves have an average duration of just $30\,{\rm min}$ (the pulsations periods are just a few min) and have often been detrended to filter out long-period variations (some binaries have been missed because of that). This dataset is therefore not suited for our purposes.

Light curves of unprecedented quality and duration have been obtained by the Kepler space mission. However, due to the pre-selection of potential planet host stars only 29 single-lined sdBs and sdOBs have been observed, among them one HW\,Vir (2M\,1938+4603) and three reflection effect systems \citep{oestensen:kepler}. This fraction of 14\% seems to be consistent with our result. However, the Kepler targets have not been selected based on RV variations and the number of spectroscopically confirmed close binaries is significantly smaller. The Kepler fraction of reflection effect and HW\,Vir systems is therefore even higher than the fraction derived here.

It has to be pointed out that our sample as well as the Kepler sample are still small and the derived number fractions therefore affected by small number statistics. Larger and more homogeneous samples like the ones from the Kepler K2 and the proposed TESS mission will shed more light on this important question.

\section{Conclusions}
\label{discussion}

We found another eclipsing sdB binary J1920 with a low-mass M dwarf companion increasing the number of known systems to 20. Moreover, a known sdB binary was found to contain a long-period pulsating sdB star.

Additionally we tried to constrain the nature of the companions for the other observed sdB binaries and the fractions of cool, low-mass and brown dwarf companions. More than $15\%$ of the close sdB binaries have cool, low-mass companions and more than $8\%$ of the sdB binaries seem to have substellar companions. This fraction is much lower than predict by \citet{han:2003}, however we are only sensitive to short periods and unseen companions like very low-mass stars or substellar objects and white dwarfs, which means our fraction is just a lower limit.

Several studies have shown that about one third of the sdBs have composite spectra and are therefore members of wide binary systems with solar-type main sequence companions. About half of the single-lined sdBs show significant RV variations on short time scales \citep[see][for a review]{heber:2016}. Taking those fractions into account we derive a close brown dwarf fraction of about $3\%$ around sdBs. 

The frequency of close brown dwarfs around sun-like stars, the likely progenitors of most sdBs, is $<1\%$ and therefore likely lower \citep{sahlmann:2011}.
This higher fraction
makes it more likely that substellar companions, which were destroyed during the common-envelope phase could also be responsible for the formation of single sdB stars \citep{geier,vs:2014_I}.

Only eight close PCEB WD+BD systems are known in double-lined binaries, as the WD is much smaller and hence fainter than an sdB star and the BD particulary in the infrared of comparable brightness. Although no reliable number fractions are known yet, the fraction of close substellar companions to the more than $40\,000$ spectroscopically identified WDs  is likely to be significantly smaller than the one derived for hot subdwarfs. This alone is a strong indication that common envelope interactions with substellar objects are preferentially followed by a hot subdwarf phase. Having a closer look at the WD masses of the eight known WD+BD system WD\,0137-349B \citep[$0.39\,M_{\rm \odot}$,][]{maxted:2006}, GD\,1400 \citep[$0.67\,M_{\rm \odot}$,][]{farihi:2004}, WD\,0837+185 \citep[$0.798\,M_{\rm \odot}$,][]{casewell:2012}, NLTT\,5306 \citep[$0.44\,M_{\rm \odot}$][]{steele:2013}, CSS\,21055 \citep[$0.53\,M_{\rm \odot}$,][]{littlefair:2014}, SDSS J155720.77+091624.6 \citep[$0.447\,M_{\rm \odot}$,][]{farihi:2017}, SDSS J1205-0242 \citep[$0.39\,M_{\rm \odot}$,][]{parsons:2017} and SDSS J1231+0041 \citep[$0.56\,M_{\rm \odot}$,][]{parsons:2017}  this conclusion is strengthened even further. Six of the known close WD+BD binaries ($75\%$) have masses between $0.39-0.56\,M_{\rm \odot}$, which is very close to the typical mass of core helium-burning hot subdwarfs. The WD mass distribution shows a strong peak at about $0.6\,M_{\rm \odot}$, while WDs with masses around $0.5\,M_{\rm \odot}$ are very rare \citep[e.g.,][]{tremblay:2009}. It is therefore  possible that those six systems evolved through an sdB phase before. 

To end up on the extreme horizontal branch, the hydrogen envelope of the progenitor star must be removed at the same time as the core of the red giant ignites helium-burning. This happens right at the tip of the red giant branch, where the star reaches its maximum radius and the envelope is most loosely bound. Maybe those conditions are favorable for very low-mass companions to eject the envelope. We conclude that sdB+BD binaries are a likely outcome of the interaction between stars and substellar objects with a suitable initial separation and therefore very important to study such interactions.


\begin{acknowledgements}
Based on observations collected at the Centro Astronómico Hispano Alemán (CAHA), operated jointly by the Max-Planck Institut für Astronomie and the Instituto de Astrofisica de Andalucia (CSIC) with BUSCA at 2.2m telescope and TWIN at the 3.5 m telescope. Also based on observations with the ISIS spectrograph at the William Herschel Telescope located at the Roques de los Muchachos observatory at La Palma, Canary Islands.\\
V.S. acknowledges funding by the Deutsches Zentrum f\"ur Luft- und Raumfahrt (grant 50 OR 1110), the Erika-Giehrl-Stiftung, and the Deutsche Forschungsgemeinschaft under grant GE\,2506/9-1. S.G. acknowledges funding by the Deutsche Forschungsgemeinschaft under grant GE\,2506/8-1.
\end{acknowledgements}

\bibliography{aabib}
\bibliographystyle{aa}
\Online

\appendix

\section{Estimation of the fraction of low-mass stellar and substellar companions}
\label{method}

\begin{figure}[h!]
	\centering
	\includegraphics[width=1.0\linewidth]{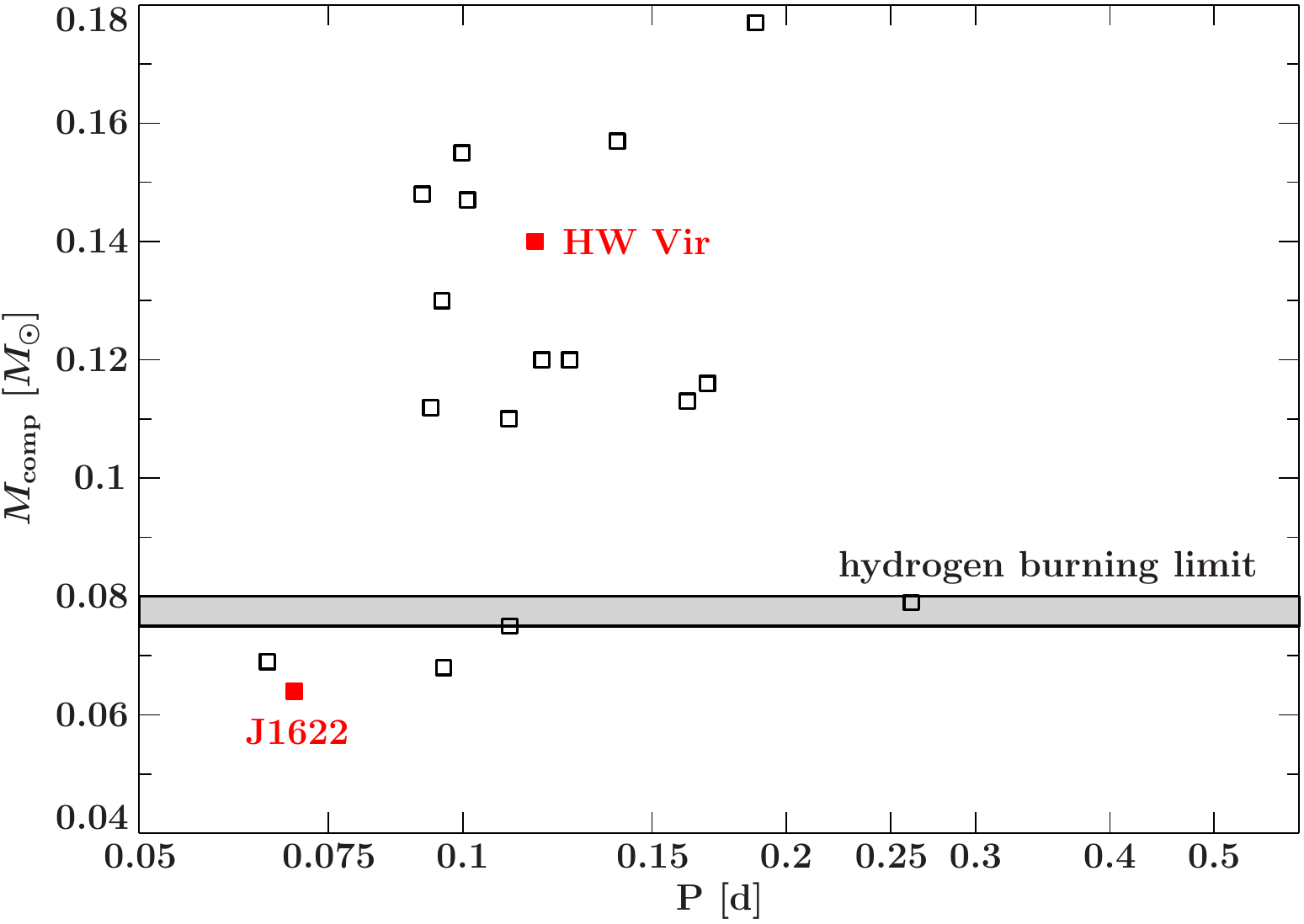}
	\caption{Period-companion mass relation of the known HW Vir systems. Our prototype systems (HW Vir for an HW Vir system with low-mass main sequence companion and J1622 for an HW Vir with brown dwarf companion) are marked in red. The parameters of the known HW Vir systems are summarized in Table \ref{HW Vir}.}
	\label{logP}
\end{figure}

\begin{figure*}
	\includegraphics[angle=-90,width=0.5\linewidth]{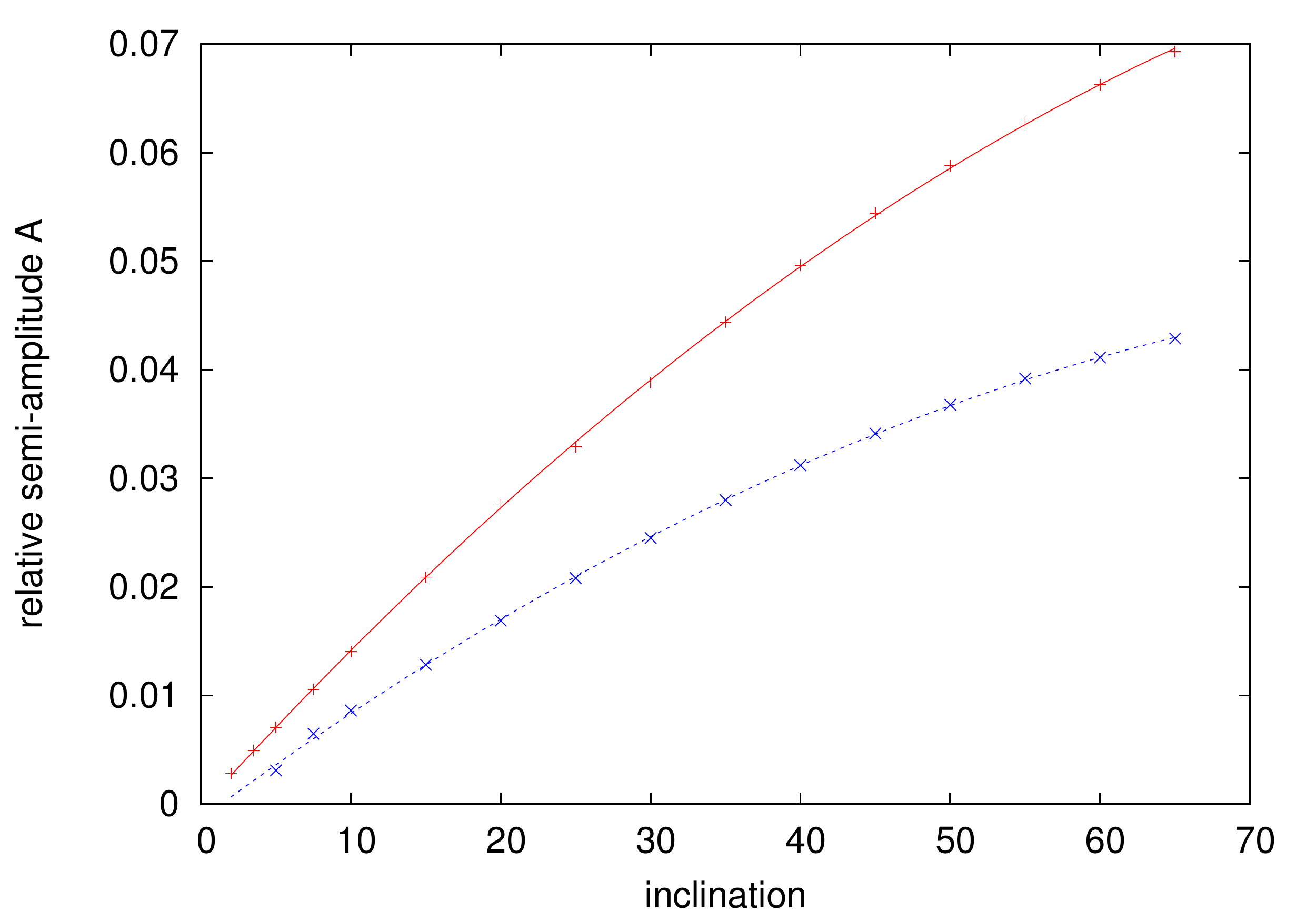}
	\includegraphics[angle=-90,width=0.5\linewidth{}]{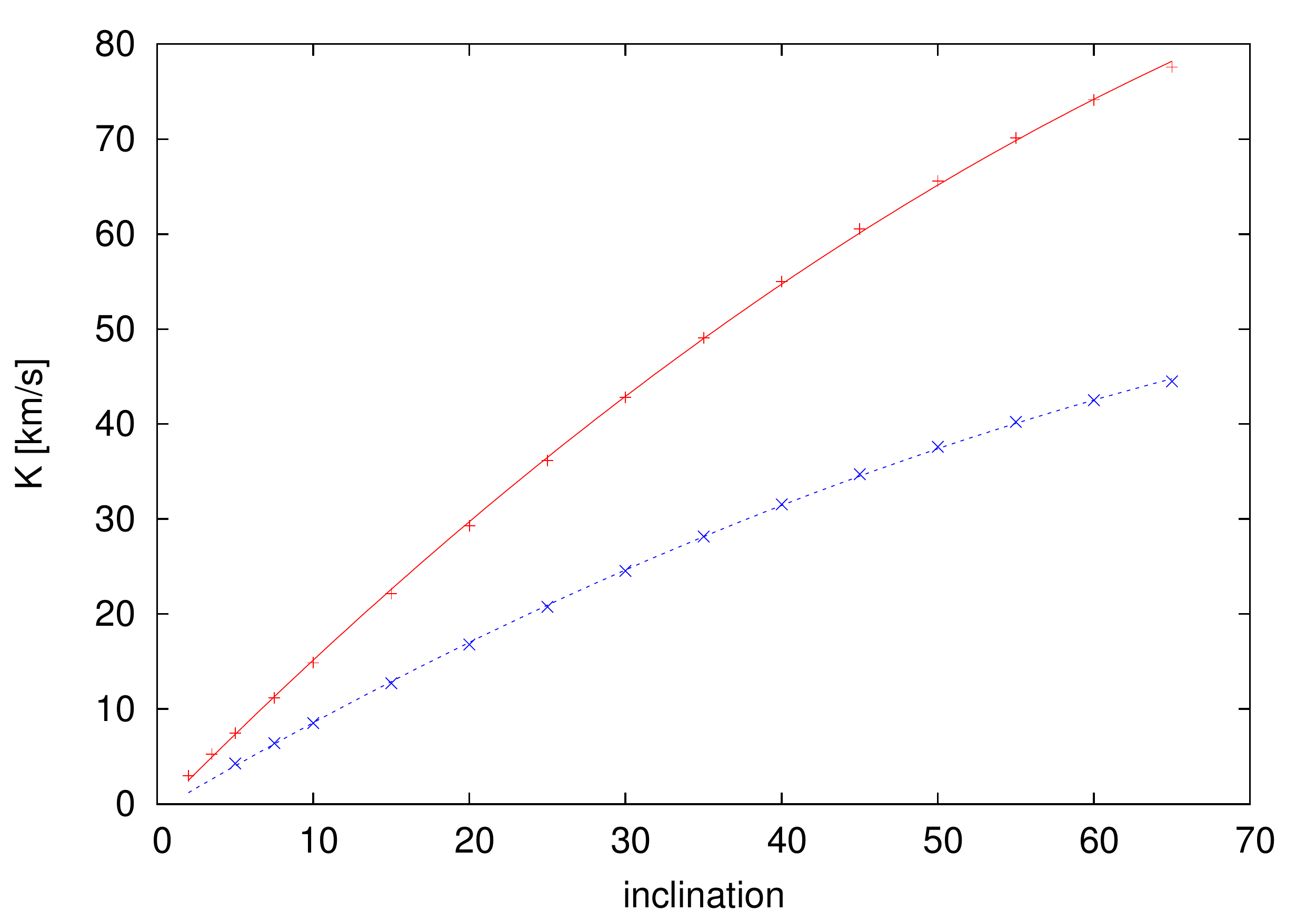}
	\caption{Expected relative semi-amplitude of the reflection effect in $R_{\rm B}$ (left panel) and expected RV semi-amplitude of the hot subdwarf (right panel) for different inclinations. They were calculated with Eq. \eqref{e1} and \eqref{e2} for the parameters of a typical HW Vir system like the prototype HW Vir (+) and for an sdB+BD system like J1622 (x).}
	\label{incl}
\end{figure*}

To estimate the true fraction of sdB binaries with cool, low-mass companions in our sample, we not only have to count the three HW\,Virs and one pulsating, reflection effect binary we found (see Sect. \ref{observ}), but also have to quantify how many similar systems (if any) remained undetected in the other 22 binary candidates, which do not show any significant variations in their lightcurves. 
Since the targets were selected spectroscopically we not only have the lightcurves but also RV shifts measured for all of them. Here we use both to put constraints on the companion type.  

In Fig. \ref{logP} we show the relation between the companion mass and the orbital period for the HW Vir systems. From this plot it can be seen that HW Vir represents a typical HW Vir system. Furthermore, J1622 is a good representative for an HW Vir system with brown dwarf companion. Hence, we choose this systems as prototype systems for the further investigation.
We calculated synthetic lightcurves adopting the parameters of HW Vir and J1622 (Table \ref{HW Vir}) varying the inclinations with the MORO code \citep{drechsel:1995}. For any given inclination angle we then derived the amplitude of the light variation. In addition, we calculated the RV amplitudes of those systems for different inclination angles. Figure \ref{incl} shows the expected semi-amplitudes of the light variations in $R_{\rm B}$ as well as the expected semi-amplitudes of the radial velocity variations for different inclinations. By fitting parabolas to the points we derived the following relations between the inclination $i$ and the relative semi-amplitude of the light variation in the $R_B$ band (650 nm) $A$ as well as the semi-amplitude of the radial velocity $K$, thereby $i$ is given in degrees. $K$ and $A$ are higher in the case of HW Vir despite the longer period of this binary with respect to J1622 due to the higher mass and larger radius of the companion:

\setlength{\abovedisplayskip}{0pt}
\setlength{\belowdisplayskip}{0pt}

\begin{subequations}
\label{e1}
\begin{equation*}
\mbox{\bf for HW Vir:}
\end{equation*}
\begin{equation}
K=-0.00691662\,i^2+1.66477\,i-0.80223
\label{e1a}
\end{equation}
\begin{equation}
A=-6.79716\cdot10^{-6}\,i^2+0.00151838\,i-0.000361173
\label{e1b}
\end{equation}
\end{subequations}

\setlength{\abovedisplayskip}{0pt}
\setlength{\belowdisplayskip}{0pt}

\begin{subequations}
\label{e2}
\begin{equation*}
\mbox{\bf for J1622:}
\end{equation*}
\begin{equation}
K=-0.00416351\,i^2+0.970709\,i-0.73049
\label{e2a}
\end{equation}
\begin{equation}
A=-5.25819\cdot10^{-6}\,i^2+0.00102431\,i-0.00137412
\label{e2b}
\end{equation}
\end{subequations}
\\

Comparing the two panels of Figure \ref{incl} it becomes clear that a close sdB+dM/BD binary seen under a small inclination angle will show small light variations and at the same time small RV variations. If a binary candidate with small or undetected light variations should show high RV variations, it cannot be a cool low-mass object, but must be a compact companion like a WD. In this case we can firmly exclude a close sdB+dM/BD binary. We will now apply this criterion to our sample.

To derive upper limits for possible, but undetected light variations we normalized the lightcurves in the $R_{\rm B}$-band and calculated mean values and standard deviations. These standard deviations are also a measure of the typical photometric uncertainties for each lightcurve. Some of the lightcurves show a linear trend with changing airmass. This is due to the fact that the target sdB star is usually much bluer than the available comparison stars and that the atmospheric extinction is wavelength dependent. We refrained from correcting for this trend, which leads to an overestimation of the standard deviation and, hence, a more conservative uncertainty estimate.

The derived standard deviations are now adopted as upper limits for the semi-amplitudes $A$ of hidden reflection effects ($A\leqq\sigma$). Using Eq. \eqref{e1b} and Eq. \eqref{e2b} we then derive the  maximum possible inclinations for hidden reflection effect systems similar to HW\,Vir and J1622 (using $A_{\max}(i_{\rm max})=\sigma$). These inclinations $i_{\rm max}$ are then converted to upper limits for the corresponding RV semi-amplitudes $K_{\rm max}(i_{\rm max})$ and RV shifts $2\cdot K_{\rm max}$ using Eq. \eqref{e1a} and \eqref{e2a}. If the measured radial velocity shift is higher than the expected one ($\Delta v_{\rm rad,max}>2\cdot K_{\rm max}$), we can exclude systems with parameters similar to HW Vir or J1622.

The results for all non-detection lightcurves are displayed in Table \ref{mass}. The table is sorted by the false detection probability for RV variability $\ln{p}$ \citep{geier:2015}. We divided the table into solved systems with known orbital parameters in bold face, RV variable systems with $\ln{p}<-9.2$ which corresponds to a false-alarm probability $<0.01\%$ and RV-variable candidates with  $\ln{p}<-3.0$ ($<5\%$). Table \ref{backup} shows the results for all RV constant and back-up targets, which have been observed during the MUCHFUSS photometric follow-up.  
Moreover, we observed nine sdOs or He-sdOs selected in the MUCHFUSS project. They are summarized in Table \ref{sdo}. Two of them showed variations as presented in Sect. \ref{var_sdo}.

Due to bad weather the lightcurves of some of our targets do not have the pursued length of 1.7 hours (the period of J1622) to 2.8 hours (the period of HW Vir). However, if the lightcurve covers only parts of the orbit it depends on the phase in which the system was observed, how large the measured amplitude of a possible reflection effect is. For example if we assume only half of the orbit was observed we would only measure half of the amplitude in the worst case. This effect has to be considered to draw firm conclusions from the lightcurves. In the case of J204613-045418, we took a lightcurve of only 0.67 hours. However, \citet{geier:2011} determined the period of the system to 0.24311 d, which is much longer than the period of HW Vir or J1622. They also derived a semi-amplitude $K_1$ of $134.3 \pm 7.8\,\rm kms^{-1}$, which is much higher than the $K_1$ of $\rm 87.9 \pm 7.8\,kms^{-1}$. Therefore, we can exclude a typical HW Vir system and a substellar companion. The same is valid for the other systems with known periods (marked in bold). In the case of J153411+543345 only 70\% of the orbital period of J1622 was covered. Hence, the amplitude of a hidden reflection effect could be as high as 20.0~\permil for a system with the parameters of J1622 or 23.7 for the parameters of HW Vir. This translates in a $K_{\rm J1622,max}$ of about $19\,\rm kms^{-1}$ and $K_{\rm HW Vir,max}$ of about $25\,\rm kms^{-1}$ still much smaller than the measured RV shift. The same was done for the other systems, which were not observed long enough. The results of this analysis can be found in Table \ref{mass}.

\begin{sidewaystable*} 
\caption{Results of the analysis of the observed lightcurves without variations in the MUCHFUSS follow-up}
\label{mass}
\begin{tabular}{l|cccccccccccc}
 target & primary &$\sigma$ \footnote{standard deviation of the normalised lightcurve in $R_{\rm B}$} &$\Delta T_{\rm lc}\footnote{length of the observed lightcurve}$& $i_{\rm J1622, max}\footnote{maximum inclination}$ & $K_{\rm J1622,max}\footnote{expected semi-amplitude of the RV curve}$ & ${i_{\rm HW Vir, max}}^{\color{red}c}$ &${K_{\rm HW Vir, max}}^{\color{red}d}$&\multicolumn{3}{c}{$\rm \Delta v_{rad,max}\footnote{maximal measured radial velocity shift}$}
 & system excluded \footnote{type of system that can be excluded}&$\ln{p}$ \footnote{False detection probability of radial velocity variability \citep[taken from][]{geier:2015}; stars were classified as RV variable for $\ln{p}<-9.2$ and RV-variable candidates for $\ln{p}<-3.0$.}\\
 &&$\permil$&h&$^\circ$&$\rm km s^{-1}$&$^\circ$&$\rm km s^{-1}$&\multicolumn{3}{c}{$\rm km s^{-1}$}&d&\\\hline\hline
  \textbf{J002323-002953}\footnote{stars with known periods from time-resolved spectroscopy \citep[see][]{kupfer:2015}} & sdB+WD & 4.00 & 1.90 & 5.40 & 8.77 & 2.76 &  7.48 & 168.0 & $\pm$ & 4.0 
  &J1622 \& HW Vir &$<-680$\\
  \textbf{J095238+625818}$^{\color{red} h}$ & sdB+WD & 3.76 & 1.54 & 5.05 & 8.14 & 2.54 &  6.75 & 154.0 & $\pm$ & 8.0 
  &J1622 \& HW Vir &$<-680$\\
  \textbf{J134632+281722}$^{\color{red} h}$ & sdB+WD & 5.57 & 3.47 & 7.03 & 11.78 & 3.83 &  10.93 & 191.0 & $\pm$ & 7.0 
  &J1622 \& HW Vir&$<-680$ \\
  \textbf{J083006+475150}$^{\color{red} h}$ & sdB+WD & 6.30 & 2.14 & 7.80 & 13.18 & 4.32 &  12.53 & 164.0 & $\pm$ & 9.0 
   &J1622 \& HW Vir &$<-680$\\
  \textbf{J032138+053840}$^{\color{red} h}$ & sdB+? & 2.60 & 1.52 & 3.96 & 6.10 & 1.82 &  4.40 & 110.0 & $\pm$ & 9.0 
  & J1622 \& HW Vir
  &$<-680$ \\  
  \textbf{J204613-045418}$^{\color{red} h}$ & sdB+WD & 5.47 & 0.67 & 6.93 & 11.59 & 3.76 &  10.71 & 259.0 & $\pm$ & 16.0 
   &J1622 \& HW Vir
  &$-480.28$ \\
  \textbf{J011857-002546}$^{\color{red} h}$ & sdOB+? & 4.39 & 1.98 & 5.80 & 9.52 & 3.02 &  8.33& 140.0 & $\pm$ & 8.0 
   & J1622 \& HW Vir
  &$-386.64$\\
\hline 
  J191908+371423 & sdB & 10.60 & 2.00 & 12.49 & 21.49 & 7.31 &  22.00 & 237.0 & $\pm$ & 12.0 
  &J1622 \& HW Vir &-526.21\\
  J072245+305233 & sdB & 7.51 & 1.50 & 9.10 & 15.51 & 5.16 & 15.19 & 123.0 & $\pm$ & 12.0 
  &J1622 \& HW Vir &-62.09\\
  J093059+025032 & sdB & 3.52 & 2.04 & 4.90 & 7.85 & 2.44 &  6.42 & 91.0 & $\pm$ & 9.0 
  &J1622 \& HW Vir&-49.22 \\
  J115358+353929 & sdOB & 16.00 & 2.18 & 18.77 & 32.05 & 11.19 &  33.92 & 79.0 & $\pm$ & 9.5 
  &J1622 \& HW Vir &-19.15\\  
  J153411+543345 & sdOB & 13.87 & 1.17 & 16.23 & 27.86 & 9.64 &  29.21 & 83.0 & $\pm$ & 18.5 
  &J1622 \& HW Vir &-12.52\\
  J224518+220746 & sdB & 5.73 & 1.93 & 7.20 & 12.09 & 3.93 &  11.28 & 70.0 & $\pm$ & 11.5 
  &J1622 \& HW Vir &-12.28\\  
  J074534+372718 & sdB & 7.45 & 2.55 & 9.03 & 15.40 & 5.12 &  15.06 & 64.0 & $\pm$ & 17.0 
  &J1622 \& HW Vir &-9.74\\
  \hline    
  J030607+382335 & sdB & 6.40 & 1.74 & 7.91 & 13.37 & 4.39 & 12.75  & 48.0 & $\pm$ & 6.5 
  &J1622 \& HW Vir&-8.85 \\ 
  J185129+182358 & sdB & 8.00 & 1.40 & 9.63 & 16.46 & 5.50 &  16.27 & 105.0 & $\pm$ & 18.0 
  &J1622 \& HW Vir &-7.33\\
  J171629+575121 & sdOB & 27.50 & 1.41 & 34.19 & 55.18 & 19.99 &  59.43 & 67.0 & $\pm$ & 15.5 
  &- &-6.14\\  
  J112242+613758 & sdB & 2.87 & 1.50 & 4.24 & 6.61 & 2.00 &  5.00 & 83.0 & $\pm$ & 18.5 
  &J1622 \& HW Vir &-5.80\\  
  J065044+383133 & sdOB & 10.60 & 2.33 & 12.49 & 21.49 & 7.31 &  22.00 & 88.0 & $\pm$ & 13.5 
  &J1622 \& HW Vir &-4.63\\  
  J115716+612410 & sdB & 5.19 & 2.17 & 6.63 & 11.05 & 3.57 &  10.09 & 102.0 & $\pm$ & 27.0 
   &J1622 \& HW Vir &-3.63\\
  J113304+290221 & sdB & 13.60 & 2.84 & 15.92 & 27.33 & 9.45 &  28.62 & 95.0 & $\pm$ & 30.0 
  &J1622 \& HW Vir  &-3.39\\
  J133639+111948 & sdB & 9.50 & 1.97 & 11.27 & 19.36 & 6.54 &  19.58 & 48.0 & $\pm$ & 14.0 
  &J1622 \& HW Vir &-3.25\\
 
\hline\hline

 \end{tabular}
 \end{sidewaystable*}

\clearpage
\section{Back-up targets}
\begin{table}[h]\centering
	\caption{Observations and light variations of the observed back-up targets.}
	\label{backup}
	\begin{tabular}{cccc}\hline\hline
		target & primary &$\sigma^a$ &$\Delta T_{\rm lc}^b$\\
		& & \permil& h\\\hline
		J074508+381106$^{a}$& sdB  & 8.30 & 0.43 \\
		J073646+220115$^{a}$ & sdB  & 27.60 & 1.72\\ 
		J234528+393505$^{a}$ & He sdO& 7.15 & 1.67\\
		J074811+435239$^{a}$ &sdB & 5.16 & 1.90\\  
		J030749+411401$^{a}$ & sdB& 16.68 & 1.74\\
		J015026-094227$^{a}$ & sdB & 6.957 & 2.21\\
		J074551+170600$^{a}$ & sdOB & 5.30 & 3.07\\ 
		J215053+131650$^{a}$ & sdB & 6.300 & 2.23\\ 
		
		J071011+403621$^{b}$ & sdB  & 17.00 & 1.42 \\
		
		HE\,2208+0126$^{c}$ &sdB & 5.64 & 1.26\\
		PG\,0026+136$^{c}$ &sdB & 5.00 & 2.61\\
		
		J121150+143716$^{d}$ & sdB  &8.50 & 2.02\\
		
		J050735+034815$^{e}$ & sdB  &3.10 & 1.61\\
		
		J092520+470330 & sdB & 8.07 & 1.87 \\
		J075937+541022 & sdB & 13.60 & 2.34 \\  
		J130439+312904 & sdOB & 7.30 & 2.12 \\
		J073701+225637 & sdB & 8.00 & 2.19\\ 
		J220810+115913 & sdB & 5.79 & 2.12 \\ 
		J052544+630726 & sdOB & 9.20 & 1.82  \\ 
		J233406+462249 & sdB & 8.15 & 1.69\\

		\hline
	\end{tabular}
	\tablefoot{\\
		$^{a}$ \citet{geier:2015}\\
		$^{b}$ {\O}stensen 2006 (Subdwarf database)\\
		$^{c}$ \citet{lisker}\\
		$^{d}$ \citet{nemeth:2016}\\
		$^{e}$ \citet{kawka:2015}\\
	}
\end{table}
\begin{table}[h]\centering
	\caption{Observations and light variations of He sdO and sdO from the MUCHFUSS project.}
	\label{sdo}
	\begin{tabular}{ccccc}\hline\hline
			target & primary &$\sigma^a$ &$\Delta T_{\rm lc}^b$\\
		& & \permil& h\\\hline
		
		J141549+111213 & He-sdO & 55.70 & 1.05\\
		J232757+483755 & He-sdO & 7.65 & 1.53\\
		J221920+394603 & sdO & 6.28 & 0.92\\
		J090957+622927 & sdO & 24.00 & 2.03 \\
		103549+092551 & He-sdO & 7.71 & 2.38 \\
		J161015+045051 & He-sdO &12.00 & 1.70\\
		J112414+402637 & He-sdO & 13.50 & 2.78\\ 
		J163416+22114 & He-sdOB & 4.34 & 2.14 \\
		J012739+404357 & sdO & 10.00 & 2.91 \\
		\hline
	\end{tabular}
\end{table}

\section{Parameters of the known HW Vir systems}
\clearpage

\begin{sidewaystable}
	\vspace{10cm}
	\tiny
	\caption{Parameters of the known HW Vir systems}
	\label{HW Vir}
	\setlength{\tabcolsep}{3.25pt}
	\renewcommand{\arraystretch}{0.9}
	\begin{tabular}{cccccccccccc}
		\toprule
		system&V&$P$	&	$K_1$&	$i$&	$a$&	$M_{\rm sdB}$&	$M_{\rm comp}$&	$T_{\rm eff}$&	$\log{g}$&$\log{y}$&reference\\
		&[mag]&[d]&$\rm km\,s^{-1}$&[$^\circ$]&[$R_{\odot}$]&[$M_{\odot}$]&[$M_{\odot}$]&[$10^3$ K]&[cgs]\\\midrule\midrule
		AA Dor&11.14&0.2615397363(4)	&$40.15 \pm0.11$	&$89.21 \pm 0.30$	&$1.4102^{+0.0863}_{-0.0341}$	&$0.4705^{+0.0975}_{-0.0354}$	& $0.0811^{+0.0184}_{-0.0102}$	&$42.0\pm1.0$& $5.46\pm0.05$ &$-2.57\pm0.07$& \citet{hoyer:2015}\\
		&&&&&&&&&&&and references therein\\
		HW Vir&10.58& 0.116719555(2) &  	$87.9\pm1.1$&	$80.98\pm0.1$&$0.860 \pm 0.01$&	$0.485 \pm 0.013$&	$0.142\pm0.004$&	$28.0\pm0.05$& $5.51\pm0.006$&$-2.45\pm0.02$&\citet{tuc_maya},\\
		&&&&&&&&&&&\citet{beuermann:2012}\\
		&&&&&&&&&&&and references therein\\
		HS2231+2441&14.1&0.11058798(8) &	$49.1\pm 3.1$&	79.6&	0.80&	0.47&	0.075&	$28.37\pm0.08$& $5.39\pm0.011$ &$-2.91 \pm 0.04$&\citet{oestensen:2007}	\\
		&&&&&	0.66&	$0.265\pm 0.01$&	0.05&	&&& \citet{oestenson:2008}\\
		NSVS 14256825&13.9&0.110374230(2)& $73.4 \pm 2.0$	&	$82.5\pm0.4$&	$0.80\pm0.04$&	$0.41\pm0.07$&	$0.109\pm0.023$&	$40.0\pm0.5$ &$5.5 \pm 0.05$ &$-2.52\pm0.12$&\citet{almeida:2012}\\
		&&&&&&$0.346 \pm 0.079$ & $0.097 \pm 0.028$&&&&\citet{almeida:2012}\\
		2M 1533+3759&13.0&0.16177042(1)	&$71.1 \pm 1.0 $	&$86.6\pm0.2$&	$0.98\pm0.04$&$0.376\pm0.055$	&	$0.113\pm0.017$&	$29.23\pm0.13$& $5.58\pm0.03$ &$-2.37 \pm 0.05$&\citet{for:2010}\\
		SDSS J162256.66+473051.1&$\sim$16.2&0.0696859(53)	&$47.2\pm2.0$	&$72.33\pm1.11$&	$0.58\pm0.02$&	$0.48\pm0.03$&	$0.064\pm0.004$&	$29.000\pm0.6$ &$5.65\pm0.06$&$-1.87 \pm 0.05$&\citet{vs:2014_I} \\
		SDSS J0820+0008&15.17&0.096240737(2)	&	$47.4\pm1.9$	&$85.83\pm0.19$	&$1.1\pm0.04$&	$0.47$&	$0.068\pm0.003$&	$26.7\pm0.1$& $5.48\pm 0.12$&$-2.0 \pm 0.07$&\citet{geier},\\
		&&&&&$0.60\pm0.02$&0.25&$0.045^{+0.003}_{0.002}$&&&&\citet{vs_diss}\\
		ASAS 10232&11.7&0.13926940(4)&	$81.0\pm3.0$	&$65.9\pm0.7$	&$0.963 \pm 0.036$	&$0.461 \pm 0.051$	&$0.157\pm0.017$&	$28.40\pm0.5$& $5.60\pm0.05$&$-1.80\pm0.2$&\citet{vs}\\ 
		SDSS J192059+372220&15.7&0.168876(35)&	$59.8\pm2.5$&	67.2&	$1.078\pm0.045$&	0.47:&	$0.116\pm0.007$&	$27.60\pm0.6$& $5.4\pm0.1$&$-2.50 \pm 0.25$& this work\\
		EC 10246-2707&14.38&0.118 507 9936(9)	&$71.6\pm1.6$&	$79.75\pm0.13$&	$0.84\pm0.1$&	$0.45\pm 0.17$&	$0.12\pm0.05$&	$28.90\pm0.5$& $5.64\pm0.06$&$-2.5\pm0.2$&\citet{barlow:2012}\\
		HS0705+6700&14.9&0.09564665(39)&	$85.8\pm3.7$&	$84.4\pm0.3$&	0.81&	0.48&	0.13&	$28.8\pm0.28$& $5.4\pm0.04$&$-2.68 \pm 0.05$& \citet{drechsel:2001}\\
		BUL--SC16 335& 16.4&0.125050278	& $92.5\pm6.2$&	$74.6\pm0.9$&	-	&0.47:	&$0.16\pm0.05$	&$31.500\pm1.8$ &$5.70\pm0.2$&$-1.8\pm0.1$&\citet{polubek:2007},\\
		&&&&&&&&&&&\citet{geier:2014}\\
		VSX J075328.9+722424&$\sim$15.7& 0.2082535(6)&&$88.215 \pm0.048$&&&&&&&\citet{pribulla:2013}\\
		Konkoly J064029.1+385652.2&$\sim$ 17.0&0.187284762(81)&$80 \pm 10$ &$87.11 \pm 0.03$&$1.249 \pm 0.010$&$0.567 \pm 0.138$&$ 0.177 \pm 0.051$&$55.0\pm3.0$&$5.97\pm0.3$&-$2.24 \pm 0.4$&\citet{derekas} \\
		NY Vir&13.7& 0.101015999(2)&	$78.6\pm0.6$	&$80.67\pm0.06$&	$0.767\pm0.005$&	$0.471\pm0.006$	&$0.123\pm0.001$&	$32.8\pm0.08$& $5.77\pm0.02$&$-2.92 \pm 0.09$ &\citet{vangrootel},\\
		&&&&&&&&&&&\citet{nyvir}\\
		2M1938+4603&12.1 &0.125765300&	$65.7\pm0.6$&	$69.45\pm0.2$&	0.89&	$0.48\pm0.03$&	$0.12\pm0.01$&	$29.564\pm0.1$& $5.425\pm0.009$&$-2.36 \pm 0.06$ &\citet{oestenson:2010}\\
		PTF1 J072455.75+125300.3&$\sim$17.4&0.09980(25)&		$95.8\pm8.2$&	$83.56\pm0.3$&	$0.766\pm0.04$&$	0.47:$&	$0.155\pm 0.02$&	$33.9\pm0.35$& $5.74\pm0.08$&$-2.02 \pm 0.07$&\citet{schindewolf}\\
		V2008&16.8&0.065817(1)	&$54.6\pm2.4$&	$86.83\pm0.45$&	$0.56\pm0.02$&	$0.47\pm0.03$&	$0.069\pm0.005$& 	$32.8\pm0.75$& $5.83\pm 0.05$&$-2.27 \pm 0.13$&\citet{vs:2015a}\\
		OGLE-GD-ECL-11388   & &            0.147806180(7)    && $81.9\pm0.1$               &&&&&&&                                                                                  \citet{hong:2017}\\
		PTF1J011339.09+225739.1  &  16.5 &   0.0933731(3) & $74.2 \pm 1.7$  &   $79.88\pm0.13$ &  $0.7226 \pm 0.0018$& 0.47: &          $0.1119 \pm 0.0030$ &$29.28 \pm 0.73$  &$5.78 \pm 0.09$ &    $-2.57 \pm 0.09$  &  \citet{wolz}\\
				\bottomrule
	\end{tabular}
\end{sidewaystable}\clearpage

\end{document}